\documentclass[aps,twocolumn,prc,floatfix,showpacs,preprintnumbers,amsmath,amssymb,
nofootinbib,groupedaddress]{revtex4-1}
\usepackage{wasysym}
\usepackage{color}
\usepackage{graphicx}
\usepackage{dcolumn}    
\usepackage{multirow}
\usepackage{bm}         
\usepackage{sidecap}
\usepackage{ulem}
\begin{document}
%
\title{Electric Dipole Polarizability in ${}^{208}$Pb: insights from the Droplet Model}

\author{X. Roca-Maza}
\email{xavier.roca.maza@mi.infn.it}
\affiliation{Dipartimento di Fisica, Universit\`a degli Studi di Milano and INFN,  Sezione di Milano, 20133 Milano, Italy}
\author{M. Centelles}
\affiliation{Departament d'Estructura i Constituents de la Mat\`eria and Institut de Ci\`encies del Cosmos (ICC), Facultat de F\'{\i}sica, Universitat de Barcelona, Diagonal {\sl 645}, {\sl 08028} Barcelona, Spain.}
\author{M. Brenna}
\affiliation{Dipartimento di Fisica, Universit\`a degli Studi di Milano and INFN,  Sezione di Milano, 20133 Milano, Italy}
\author{X. Vi\~nas}
\affiliation{Departament d'Estructura i Constituents de la Mat\`eria and Institut de Ci\`encies del Cosmos (ICC), Facultat de F\'{\i}sica, Universitat de Barcelona, Diagonal {\sl 645}, {\sl 08028} Barcelona, Spain.}
\author{G. Col\`o}
\affiliation{Dipartimento di Fisica, Universit\`a degli Studi di Milano and INFN,  Sezione di Milano, 20133 Milano, Italy}
\author{B. K. Agrawal}
\affiliation{Saha Institute of Nuclear Physics, Kolkata 700064, India}
\author{N. Paar}
\affiliation{Physics Department, Faculty of Science, University of Zagreb, Zagreb, Croatia}
\author{J. Piekarewicz}
\affiliation{Department of Physics, Florida State University, Tallahassee, FL 32306, USA}
\author{D. Vretenar}
\affiliation{Physics Department, Faculty of Science, University of Zagreb, Zagreb, Croatia}

\date{\today} 

\begin{abstract}
We study the electric dipole polarizability $\alpha_D$ in ${}^{208}$Pb
based on the predictions of a large and representative set of
relativistic and non-relativistic nuclear mean field models. We adopt
the droplet model as a guide to better understand the
correlations between $\alpha_D$ and other isovector
observables. Insights from the droplet model suggest that the
product of $\alpha_D$ and the nuclear symmetry energy at
saturation density $J$ is much better correlated with the neutron 
skin thickness $\Delta r_{np}$ of ${}^{208}$Pb than the polarizability
alone. Correlations of $\alpha_D J$ with $\Delta r_{np}$ and 
with the slope of the symmetry energy $L$ suggest that 
$\alpha_D J$ is a strong isovector indicator. Hence, we explore 
the possibility of constraining the isovector sector of the
nuclear energy density functional by comparing our theoretical
predictions against measurements of both $\alpha_D$ and the 
parity-violating asymmetry in ${}^{208}$Pb.
We find that the recent experimental determination of $\alpha_D$
in ${}^{208}$Pb in combination with the range for the 
symmetry energy at saturation density $J\!=\![31\pm (2)_{\rm est.}]$\,MeV suggests 
$\Delta r_{np}({}^{208}{\rm Pb}) = 0.165 \pm (0.009)_{\rm exp.}
\pm (0.013)_{\rm theo.} \pm (0.021)_{\rm est.} {\rm fm}$ and
$L= 43 \pm(6)_{\rm exp.} \pm (8)_{\rm theo.}\pm(12)_{\rm est.}$ MeV.
\end{abstract}


\maketitle 


\section{Introduction}
\label{introduction}

Experimental and theoretical studies of isospin sensitive observables, 
such as the electric dipole polarizability, the neutron skin thickness, 
and the parity violating asymmetry, are crucial for a better understanding 
of the isovector sector of the nucleon-nucleon effective interaction 
and for constraining present and future nuclear energy density 
functionals (EDFs)\,\cite{tamii11,prex,piekarewicz12}. The isovector 
properties of the nuclear Equation of State (EoS) are governed by the 
nuclear symmetry energy. The symmetry energy $S(\rho)$ encodes the 
energy cost per nucleon in converting all the protons into neutrons 
in symmetric nuclear matter. Knowledge of the symmetry energy and of 
its density dependence is critical for understanding many properties 
of a variety of nuclear and astrophysical systems, such as the 
ground and excited state properties of nuclei\,\cite{tsang12}, many 
aspects of heavy-ion collisions at different projectile-target 
asymmetries\,\cite{li2008}, and the structure, composition, and dynamics 
of neutron stars\,\cite{lattimer07}.

The electric dipole polarizability $\alpha_D$ in ${}^{208}$Pb has been
recently measured at the Research Center for Nuclear Physics (RCNP)
\cite{tamii11} via polarized proton inelastic scattering at forward
angles. This experimental technique allows the extraction of the
electric dipole response in ${}^{208}$Pb over a wide energy range with
high resolution\,\cite{tamii11}. By taking the average of all available 
data on the electric dipole polarizability in 
${}^{208}$Pb\,\cite{schelhaas88,veyssiere70}, a value of 
$\alpha_D$ = 20.1$\pm$0.6 fm$^3$ was reported\,\cite{tamii11}. 
This value, in combination with the covariance analysis performed 
for a given Skyrme functional \cite{reinhard10} constrained the 
neutron skin thickness in ${}^{208}$Pb to be 
$\Delta r_{np} = 0.156^{+0.025}_{-0.021}$ fm \cite{tamii11}. 
A subsequent systematic study based on a large class of EDFs was able to
confirm the correlation between $\alpha_D$ and $\Delta r_{np}$ 
\cite{piekarewicz12}. This study extracted a neutron skin thickness 
$\Delta r_{np} = 0.168\pm 0.022$ fm using the same experimental value 
of $\alpha_D$.
 
The purpose of this manuscript is threefold. First, we resort to a macroscopic 
approach for describing the dipole polarizability, which enables one to 
qualitatively understand, in a simple and transparent way, the correlation between
the electric dipole polarizability and the parameters that characterize the nuclear 
symmetry energy. Second, through a comprehensive ensemble of microscopic 
calculations performed with different types of EDFs\,\cite{skyrme1,*skyrme2,*lns,*sk255,roca-maza12b,*roca-maza13,sv,Lalazissis:1996rd,nl3fsu1,piekarewicz11,fattoyev13,ddme2,*ddme} we provide a quantitative analysis which allows to define the regions where 
the experiment and
the adopted microscopic approaches are compatible. Finally, the isospin properties 
of the considered EDFs are further investigated by the analysis of the dipole 
polarizability in combination with the parity violating asymmetry measured in 
polarized elastic electron scattering.

The manuscript has been organized as follows. In Sec.\,\ref{theory} we introduce
the microscopic and macroscopic models used in this work. In particular, we discuss
some of the critical insights provided by the macroscopic droplet model. In the next
section results are presented for the correlations between the electric dipole polarizability
and both the neutron skin thickness and the parity violating asymmetry. Finally, we offer
our conclusions in Sec.~\ref{conclusion}.

\section{Theoretical framework}  
\label{theory}

In the present section we introduce the theoretical formalism that will be used
to compute the various observables discussed in this work. In particular, we briefly 
review the mean-field plus Random Phase Approximation (RPA) techniques used to 
compute the distribution of isovector dipole strength. Moreover, we make connection
to the macroscopic droplet model (DM) and discuss the critical insights that emerge from 
such a simplified, yet powerful, description.

\subsection{Microscopic models}
\label{micmodel}

For the theoretical calculations presented in this work we use a set of
non-relativistic and relativistic self-consistent mean field models 
to predict ground-state properties of finite nuclei at either the 
Hartree-Fock or Hartree levels, respectively. These mean field models 
have been accurately calibrated to certain ground-state data, such 
as binding energies and charge radii of selected nuclei (including ${}^{208}$Pb) 
as well as to a few empirical properties of infinite nuclear matter at, 
or around, saturation density. To deal with dynamic properties of the system,
such as the electric dipole polarizability, the models adopt the linearization 
of the time-dependent Hartree or Hartree-Fock equations in a fully self-consistent 
manner. That is, the residual interaction employed in the calculation of
the linear response is consistent with the one used to generate the mean-field
ground state. This technique is widely known as the Random Phase 
Approximation\,\cite{ringschuck}. From the RPA calculations we obtain the 
distribution of the electric dipole strength $R(\omega;E1)$ by considering 
the dipole operator 
\begin{equation}
\mathcal{D} = \frac{Z}{A} \sum_{n=1}^N r_n Y_{1M}(\hat r_n) - \frac{N}{A}
\sum_{p=1}^Z r_p Y_{1M}(\hat r_p)\;, \label{diop} 
\end{equation}
where $N$, $Z$, and $A$ are the neutron, proton, and mass numbers,
respectively, $r_{n(p)}$ indicates the radial coordinate for neutrons
(protons), and $Y_{1M}(\hat r)$ is the corresponding spherical
harmonic. Using this definition of the dipole operator allows one to
eliminate any contamination to the physical response from the 
spurious state \cite{ringschuck,mizuyama2012}. Further details 
about these RPA calculations may be found 
in\,\cite{reinhard10,colo13,piekarewicz11,ddme2,*ddme} and references
therein. Once the electric dipole strength $R(\omega;E1)$ is determined 
as a function of the excitation energy $\omega$, the dipole polarizability
$\alpha_D$ can be computed as
\begin{equation}
\alpha_D  
= \frac{8\pi e^2}{9} \int_{0}^{\infty}\!\omega^{-1} R(\omega;E1)\,d\omega 
= \frac{8\pi e^2}{9} m_{-1}(E1) \;,
\label{alphad}
\end{equation}
where $m_{-1}(E1)$ is the sum of inverse energy weighted strength.

\subsection{Macroscopic model}
\label{macmodel}

The RPA formalism described above suggests that the extraction of the 
inverse energy weighted sum requires the evaluation of the full distribution 
of dipole strength $R(\omega;E1)$. However, given that only the $m_{-1}$
moment is required---as opposed to the full distribution of strength---a 
significantly more efficient computation of the dipole polarizability relies on 
the so-called dielectric theorem\,\cite{bohigas1979,capelli2009}. 
In this case, one solves the ground-state problem associated with
the model Hamiltonian $\mathcal{H}$ under the constraint of a weak one-body term of the
form $\lambda\mathcal{D}$, where $\mathcal{D}$ is the dipole operator.
That is, one searches for the constrained wave function $\vert\lambda\rangle$
solution of $\mathcal{H}^\prime=\mathcal{H}+\lambda\mathcal{D}$.   
The dielectric theorem establishes that the $m_{-1}$ moment may be
computed from the expectation value of the Hamiltonian in the
constrained ground state as
\begin{equation} 
m_{-1}(E1) = \frac{1}{2}\left.\frac{\partial^2
\langle \lambda \vert \mathcal{H} \vert \lambda\rangle}{\partial \lambda^2}\right\vert_{\lambda = 0} . 
\label{cons}
\end{equation}
Note that this represents an enormous simplification, as the constrained 
energy may be obtained from a mean-field calculation, without recourse to
the RPA.

Applying the same type of procedure but solving the constrained problem classically 
by using the DM approach of Myers and Swiatecki\,\cite{myers74} 
one obtains the following result:
\begin{equation} 
\alpha_D^{\rm DM} = \frac{\pi e^2}{54} \frac{A \langle
r^2\rangle}{J} \left(1+\frac{5}{3}\,\frac{9J}{4Q}A^{-1/3}\right),
\label{dpdm1} 
\end{equation} 
which was first derived by Meyer, Quentin, and
Jennings\,\cite{meyer82}. In this equation $\langle r^2\rangle$ 
is the mean-square radius of the nucleus, 
$J$ is the nuclear symmetry energy at saturation density, and $Q$ 
is the so-called surface stiffness coefficient---which measures the resistance 
of neutrons against being separated from protons\,\cite{myers74}.

It was shown in Ref.\,\cite{warda09} using a large set of EDFs that the 
ratio $J/Q$ appearing in Eq.~(\ref{dpdm1}) is linearly related to the slope of
the symmetry energy at saturation density $L$. Moreover, the DM gives the 
symmetry energy coefficient $a_{\rm sym}(A)$ of a finite nucleus of mass number 
$A$ as follows\,\cite{myers74,centelles09}:
\begin{equation}
a_{\rm sym}(A) = \frac{J}{1+\frac{9J}{4Q}A^{-1/3}}\,.
\label{asym}
\end{equation}
Expanding Eq.\,(\ref{asym}) to first order in the ``small'' parameter $J A^{-1/3}/Q$ 
[as was done in deriving Eq.\,(\ref{dpdm1})] we can write Eq.\,(\ref{dpdm1}) as
\begin{equation}
\alpha_D^{\rm DM} \approx \frac{\pi e^2}{54} \frac{A \langle r^2\rangle}{J}
\left(1+\frac{5}{3}\frac{J-a_{\rm sym}(A)}{J}\right).
\label{dpdm1asy}
\end{equation}
Given that the difference between $J$ and $a_{\rm sym}(A)$ is directly 
related to the surface symmetry energy, the above result reveals that the 
electric dipole polarizability is sensitive to the ratio of the surface and bulk 
nuclear symmetry energies\,\cite{satula06}. 

The DM may also be used to provide an expression for the neutron
skin thickness in terms of a few bulk nuclear 
properties\,\cite{myers1980,centelles09,warda09}. That is,
\begin{equation}
\Delta r_{np}^{\rm DM} = 
\sqrt{\frac{3}{5}} \left[\frac{3 r_0}{2}\, \frac{\frac{J}{Q}(I-I_C)}
{1+\frac{9J}{4Q}A^{-1/3}}\right]
 + \Delta r_{np}^{\rm coul} + \Delta r_{np}^{\rm surf}\,, \label{rnp}
\end{equation}
where $I\!\equiv\!(N\!-\!Z)/A$ is the relative neutron excess, 
$r_{0}$ is related to the saturation density $\rho_{0}$ by
$\rho_{0}\!=\! 3/(4\pi r_{0}^{3})$,
$I_C= (e^2 Z )/(20 J R)$, $R\equiv r_0 A^{1/3}$,
$\Delta r_{np}^{\rm coul}\!=\!-\sqrt{3/5}(e^2 Z)/(70 J)$ is a
correction caused by the electrostatic repulsion,
and $\Delta r_{np}^{\rm surf}\!=\!\sqrt{3/5}[5(b_n^2-b_p^2)/(2R)]$ is a 
correction caused by the difference between the surface widths
$b_n$ and $b_p$ of the neutron and proton density profiles\,\cite{myers1980,warda09}.

In this manner, one may use the DM to relate the dipole polarizability to the
neutron skin thickness. For this purpose, one expands Eq.\,(\ref{rnp}) to first-order
in $J A^{-1/3}/Q$ and finds after some algebra the following relation:
\begin{equation}
\alpha_D^{\rm DM} \!\approx\! \frac{\pi e^{2}}{54} \frac{A \langle
r^2\rangle}{J}\!\left[1\!+\!\frac{5}{2} \frac{\Delta r_{np}^{\rm DM}\!-\!
 \Delta r_{np}^{\rm coul} \!-\! \Delta r_{np}^{\rm
surf}}{\langle r^2\rangle^{1/2}(I\!-\!I_C)}\right]\,.
\label{dpdm2}
\end{equation}
Adopting a value of $J\!=\!31 \pm 2$ MeV as a reasonable
estimate compatible with recent compilations\,\cite{tsang12,lattimer12}, 
one finds for ${}^{208}$Pb that $I_{C}\!\approx\!0.028\pm0.002$
and $\Delta r_{np}^{\rm coul}\!\approx\!-0.042 \pm 0.003$ fm.
Moreover, in Ref.\,\cite{centelles10} it was shown 
that $\Delta r_{np}^{\rm surf}\!\approx\!0.09 \pm 0.01$ fm for
${}^{208}$Pb according to the predictions 
of a large sample of EDFs. Consequently, as a first reasonable
approximation, one can neglect the small variations of $I_C$, 
$\Delta r_{np}^{\rm coul}$, and $\Delta r_{np}^{\rm surf}$ in Eq.~(\ref{dpdm2})
and explicitly show that for ${}^{208}$Pb the product $\alpha_D^{\rm DM}
J$ is linearly correlated with $\Delta r_{np}^{\rm DM}$---in agreement 
with Ref.~\cite{satula06}. 

Given the well-known correlation between the neutron skin thickness 
of a heavy nucleus and the slope of the symmetry energy at saturation density 
$L\!\equiv\!3\rho_0 (dS/d\rho)_{\rho=\rho_0}$ implied by
a large set of EDFs\,\cite{brown00,*furnstahl02,roca-maza11}, one can 
also anticipate the emergence of a linear correlation between 
$\alpha_D J$ and $L$. To do so we rely on the findings of 
Ref.\,\cite{centelles09} that suggest that the symmetry energy 
coefficient of a finite nucleus is very close to that of the infinite system 
at an appropriate sub-saturation density $\rho_A$ 
[{\sl i.e.,} $a_{\rm sym}(A)\!\approx\!S(\rho_A)$]. Note that the density 
$\rho_A$ approximately obeys the following simple formula:
\begin{equation}
 \rho_A=\frac{\rho_0}{1+c A^{-1/3}} \;,
\label{roa}
\end{equation}
where $c$ can be chosen so that $\rho_{208}\!=\!0.1$\,fm${}^{-3}$. 
Using these results in Eq.~(\ref{dpdm1asy}) after expanding
$S(\rho_A)$ around saturation density, namely,
\begin{equation}
  a_{\rm sym}(A) \approx S(\rho_A) =  J -L \epsilon_A + \cdots \;,
\label{asymA}
\end{equation}
one arrives at 
\begin{equation}
\alpha_D^{\rm DM} \approx \frac{\pi e^{2}}{54} \frac{A \langle r^2\rangle}{J}
\left[1+\frac{5}{3}\frac{L}{J}\epsilon_A\right]\;.
\label{dpdm3}
\end{equation}
Note that we have defined $\epsilon_A\!=\!(\rho_0-\rho_A)/3\rho_0$,
which is approximately equal to $\epsilon_A\!=\!1/8$ for $\rho_0=0.16$ 
fm${}^{-3}$ for the case of ${}^{208}$Pb. This formula
suggests how $J$ and $L$ can be related if the dipole
polarizability is known.

\section{Results}
\label{results}

In this section we study correlations between the electric dipole polarizability 
(mostly in the form of $\alpha_{D}J$) and both the neutron skin thickness and 
parity-violating asymmetry in ${}^{208}$Pb. Our microscopic analysis involves a 
large and representative body of EDFs. We employ non-relativistic Skyrme EDFs 
widely used in the literature (labeled as Skyrme in the 
figures\,\cite{skyrme1,*skyrme2,*lns,*sk255}) and six different families
of systematically varied interactions produced, respectively, by a variation of the 
parameters around an optimal value (without significantly compromising the 
quality of the merit function). Two of the families are based on non-relativistic Skyrme 
EDFs (labeled in the figures as SAMi\,\cite{roca-maza12b,*roca-maza13} and SV\,\cite{sv}), 
while three families are based on meson-exchange covariant EDFs (labeled as 
NL3/FSU \cite{Lalazissis:1996rd,nl3fsu1,piekarewicz11}, and TF\,\cite{fattoyev13}). 
The last family is based on a meson-exchange covariant EDF but assuming 
density-dependent coupling constants (labeled as DD-ME\,\cite{ddme2,*ddme}).

\begin{figure*}[t!]
\includegraphics[width=0.45\linewidth,clip=true]{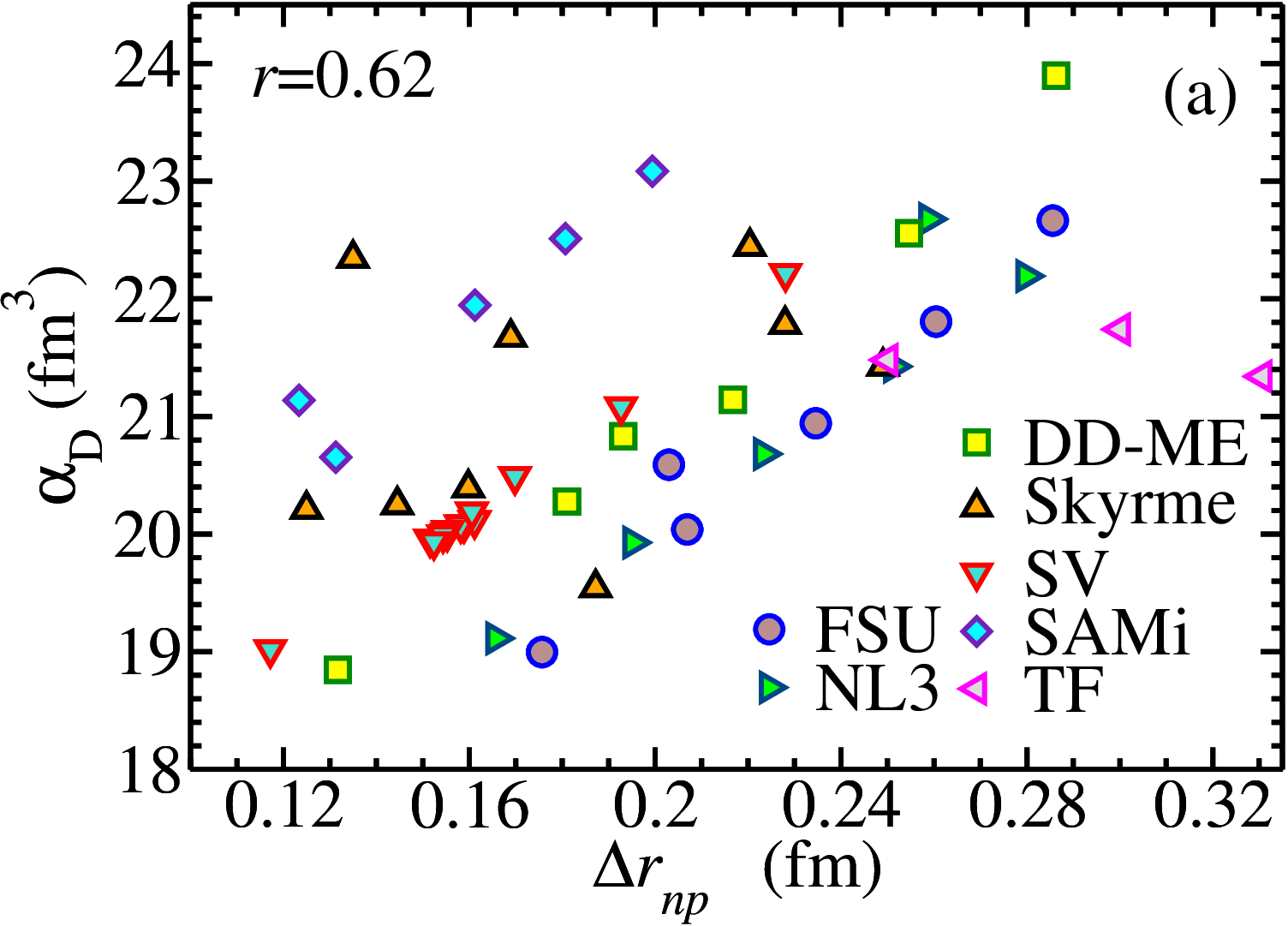}
\hfill 
\includegraphics[width=0.45\linewidth,clip=true]{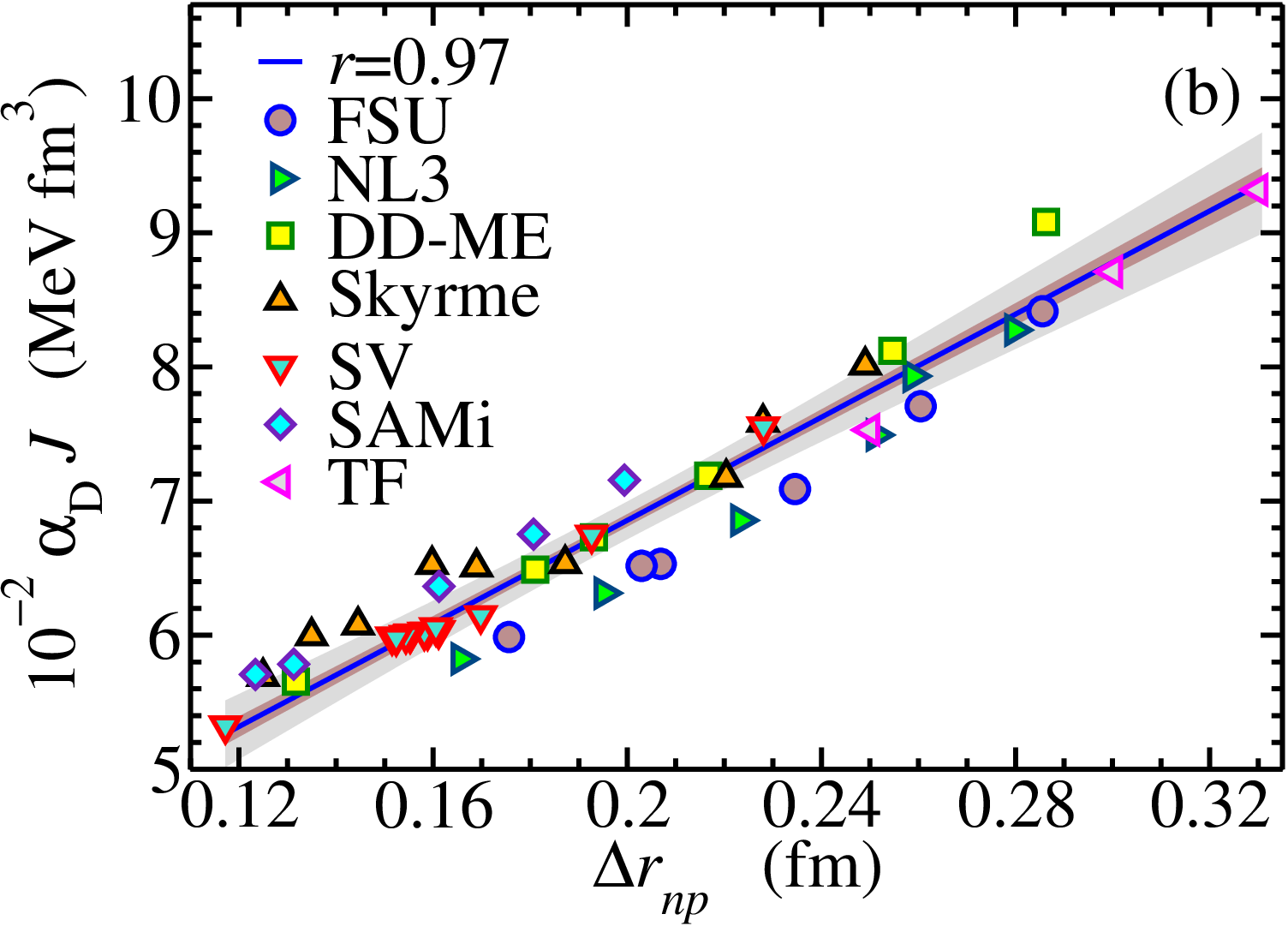} 
\caption{(Color online) (a) Dipole polarizability against
the neutron skin thickness in ${}^{208}$Pb predicted by modern nuclear
EDFs\,\cite{skyrme1,*skyrme2,*lns,*sk255,roca-maza12b,*roca-maza13,
sv,Lalazissis:1996rd,nl3fsu1,piekarewicz11,fattoyev13,ddme2,*ddme}. A
correlation coefficient of $r=0.62$ is found. 
(b) Dipole polarizability times the symmetry energy at saturation of
each model against the neutron skin thickness in ${}^{208}$Pb
predicted by the same EDFs of panel (a). The linear fit gives
$10^{-2}\alpha_D J = (3.01\pm 0.32) + (19.22\pm 0.73)\Delta r_{np}$
with a correlation coefficient $r=0.97$ and the two shaded regions
represent the 99.9\% and 70\% confidence bands.}
 \label{fig1}
\end{figure*}

\subsection{The dipole polarizability and the neutron skin thickness in ${}^{208}$Pb}  
\label{resultsa}

We start by displaying in Fig.\,\ref{fig1}(a) the dipole polarizability
$\alpha_D$ as a function of the neutron skin thickness in ${}^{208}$Pb
as predicted by the large set of EDFs employed in this work. This
figure is reminiscent of the corresponding Fig.\,1 of Ref.~\cite{piekarewicz12} 
where a significant amount of scatter between the different calculations was
observed, although a linear behavior was seen within each family of
systematically varied interactions. These trends are confirmed 
in our figure that displays a correlation coefficient of only $r\!=\!0.62$. 
Remarkably, the large spread in the model predictions is practically
eliminated by scaling the dipole polarizability by $J$ of the model.
Indeed, the microscopic calculations shown in Fig.\,\ref{fig1}(b) support the 
correlation between $\alpha_D J$ and $\Delta r_{np}$ as suggested 
by the DM approach, and clearly demonstrate---by comparing the 
two panels of Fig.~\ref{fig1}---that $\alpha_D J$ is far better correlated 
to the neutron skin thickness of ${}^{208}$Pb than the polarizability alone;
note that the correlation coefficient has increased all the way to 
$r\!=\!0.97$.  

The strength of the correlation shown in Fig.\ref{fig1}(b) allows one to reliably 
estimate, within the validity of our theoretical framework, the 
value of the neutron skin thickness of ${}^{208}$Pb as a function of 
$J$---or viceversa---once the experimental value of 
$\alpha_D$=20.1$\pm$0.6 fm$^3$\,\cite{tamii11} is assumed:
\begin{eqnarray}
     \Delta r_{np} \!= &&-0.157 \pm (0.002)_{\rm theo.} \nonumber \\
      &&+\big[1.04 \pm (0.03)_{\rm exp.}\! \pm (0.04)_{\rm theo.} \big]
     \!\times\!10^{-2} J,\quad
\label{rnpj}
\end{eqnarray} 
where $\Delta r_{np}$ is expressed in fm and $J$ in MeV. The 
``exp.'' uncertainties refer to the propagation of the experimental 
uncertainty of $\alpha_D$, whereas the ``theo.'' uncertainties
are associated to the confidence bands resulting from the linear 
fit shown in Fig.~\ref{fig1}. The theoretical uncertainties are
meant to indicate the region allowed by the employed EDFs.
Moreover, adopting $J\!=\![31 \pm (2)_{\rm est.}]$\,MeV as 
a realistic range of values for the symmetry 
energy\,\cite{tsang12,lattimer12}, and combining this estimate 
with the measured value of the dipole polarizability\,\cite{tamii11}, 
we extract from Fig.\,\ref{fig1}(b) the following constraint on
the neutron skin thickness of ${}^{208}$Pb:
\begin{equation} 
\Delta r_{np} = 0.165 \pm (0.009)_{\rm exp.}
\pm (0.013)_{\rm theo.} \pm (0.021)_{\rm est.} {\rm fm}\;.
\label{rnp0}
\end{equation} 
We have labeled the uncertainty derived from the different estimates
on $J$ as ``est.'' because it contains uncertainties coming from both
experimental and theoretical analyses which are often not easy to 
separate. In addition,  we use a different label to keep track of the 
magnitude of the various uncertainties. Finally, we note that the 
above result for the neutron skin thickness of ${}^{208}$Pb is in agreement 
with previous estimates\,
\cite{tamii11,prex,piekarewicz12,tsang12,roca-maza13,agrawal12}.

\begin{figure}[t]                    
\includegraphics[width=0.95\linewidth,clip=true]{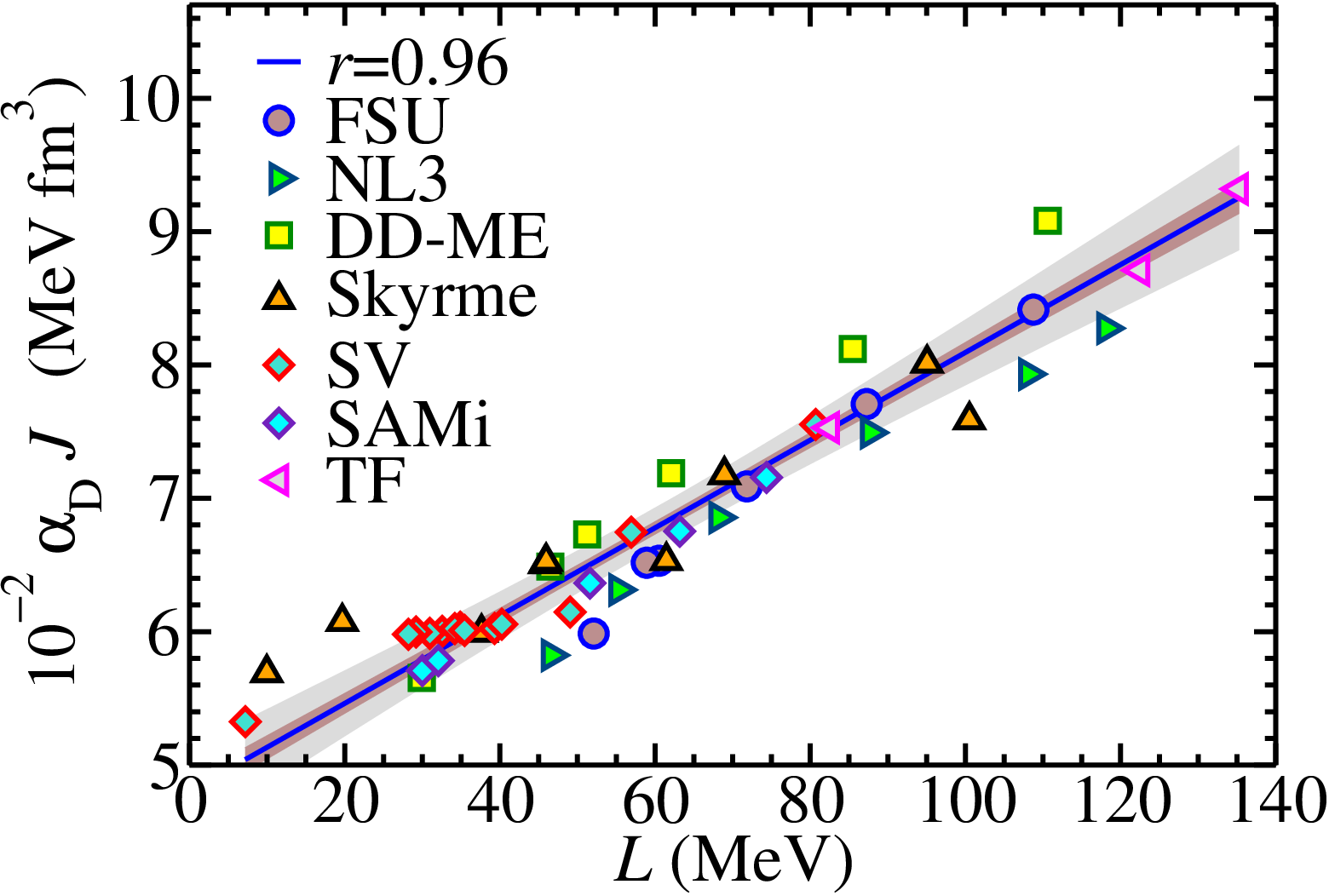}
\caption{(Color online) Dipole polarizability in ${}^{208}$Pb times
the symmetry energy at saturation as a function of the slope parameter
$L$. The same EDFs \cite{skyrme1,*skyrme2,*lns,*sk255,
roca-maza12b,*roca-maza13,sv,Lalazissis:1996rd,nl3fsu1,piekarewicz11,
fattoyev13,ddme2,*ddme} of Fig.\,\ref{fig1} are used.
The linear fit gives $10^{-2}\alpha_D J = (4.80\pm 0.04) +
(0.033\pm0.001)L$ with a correlation coefficient $r=0.96$ and the two
shaded regions represent the 99.9\% and 70\% confidence bands.}
\label{fig2}
\end{figure}

Given the strong correlation between the neutron skin-thickness of
${}^{208}$Pb and the slope of the symmetry energy $L$, one expects that
the strong correlation between $\alpha_D J$ and $\Delta r_{np}$ will
extend also to $L$. Moreover, based on the DM insights summarized
in Eq.(\ref{dpdm3}), we display in Fig.\,\ref{fig2} the microscopic
predictions for $\alpha_D J$ as a function of $L$ for the same models 
depicted in Fig.\,\ref{fig1}. The correlation between $\alpha_D J$ and 
$L$ is of particular interest since it
provides a direct relation between $J$ and $L$ via the high-precision
measurement of the electric dipole polarizability. Specifically, we
obtain
\begin{equation}
L\!=\!-146\pm(1)_{\rm theo.}
  \!+\!\big[6.11\pm(0.18)_{\rm exp.}\pm(0.26)_{\rm theo.} \big]J,
\label{jl}
\end{equation}
where both $J$ and $L$ are expressed in MeV. In particular,  adopting 
as before a value of $J\!=\![31 \pm (2)_{\rm est.}]$\,MeV, the above equation
translates into the follow constraint on~$L$:
\begin{equation}
L= 43 \pm(6)_{\rm exp.} \pm (8)_{\rm theo.}\pm(12)_{\rm est.}\;{\rm MeV}\,.
\label{l}
\end{equation}

Our results show that the analytical formulas (\ref{dpdm2}) and (\ref{dpdm3}) 
reproduce the trends of the employed microscopic models. For completeness, we now evaluate
the quantitative accuracy of these macroscopic formulas in reproducing the
present self-consistent results. In doing so, we use the microscopic
predictions for the different quantities appearing in the r.h.s. of 
Eqs.\,(\ref{dpdm2}) and (\ref{dpdm3}) and calculate $\alpha_D$ by using the two macroscopic
expressions. As a result, compared with the actual self-consistent values
of $\alpha_D$, we find that Eqs.\,(\ref{dpdm2}) and (\ref{dpdm3}) are accurate within a 10\%
and 12\% on average, respectively.

We conclude this section noting that the analysis presented here may be systematically extended to other 
heavy nuclei if $\alpha_D$ is experimentally known. This could help tighten the constraint between $J$ and $L$.

\subsection{The dipole polarizability and the parity 
violating asymmetry in ${}^{208}$Pb} 
\label{resultsb}

The parity violating asymmetry in the  elastic scattering of
high-energy polarized electrons from ${}^{208}$Pb has been recently 
measured at low momentum transfer at the Jefferson Laboratory 
by the Lead Radius Experiment (PREX) collaboration\,\cite{prex}. 
The parity violating asymmetry is defined as the relative difference 
between the differential cross sections of ultra-relativistic elastically
scattered electrons with positive and negative 
helicity\,\cite{horowitz98,*horowitz01c,*vretenar00}:
\begin{equation}
A_{\rm PV} = \Big( \frac{d\sigma_+}{d\Omega}-\frac{d\sigma_-}
{d\Omega} \Big) \bigg/ \Big( \frac{d\sigma_+}
{d\Omega}+\frac{d\sigma_-}{d\Omega} \Big) \,.
\label{apv}
\end{equation}
This landmark experiment by the PREX collaboration 
constitutes the first purely electro-weak measurement of
the neutron skin thickness of a heavy nucleus\,\cite{prex}. 
In a plane-wave Born approximation the
parity violating asymmetry is directly proportional to the
weak-charge form factor of the nucleus---itself closely
related to the neutron form factor. In exact calculations
where Coulomb distortions are taken into account a highly linear 
relation has been found between $A_{\rm PV}$ and $\Delta r_{np}$ in
${}^{208}$Pb within the realm of nuclear EDFs
 (see Fig.\,2 of Ref.~\cite{roca-maza11}). The measured 
value of the parity violating asymmetry at an average 
momentum transfer of 
$\langle Q^2\rangle\!=\!0.0088\pm 0.0001$ GeV$^2$
reported by the PREX collaboration is given by
\begin{equation}
  A_{\rm PV}=0.656\pm(0.060)_{\rm stat.} \pm (0.014)_{\rm syst.}\;{\rm ppm} \;.
\label{apv2}
\end{equation}
The experimental uncertainty of 9\% (dominated by the statistical error) 
is about three times as large as originally anticipated. By invoking some mild 
theoretical assumptions, the measurement of $A_{\rm PV}$ was used to
extract the following value of the neutron skin thickness in 
${}^{208}$Pb\,\cite{prex, horowitz12}:
\begin{equation}
 \Delta r_{np}=0.302\pm (0.175)_{\rm exp.}\pm 
 (0.026)_{\rm theo.} \pm (0.005)_{\rm strange}\;{\rm fm}\;. 
\label{rnpprex}
\end{equation}
The last contribution to the uncertainty is associated with the 
experimental uncertainty in the determination of the electric 
strange quark form factor. The result is consistent with previous 
estimates---although the central value is larger than the one
extracted from the predictions of a large set of EDFs as well
as from previous measurements of $\Delta r_{np}$ in ${}^{208}$Pb 
using hadronic probes\,\cite{tsang12}. We note, however, that 
one of the main virtues of an electro-weak extraction of $\Delta r_{np}$ 
is that it is free from most strong-interaction uncertainties.
As mentioned, the main source of the experimental uncertainty in PREX arose from 
the limited statistics, and a new run PREX-II
aiming at the original 3\% accuracy in the determination of
$A_{\rm PV}$ has been scheduled at the Jefferson 
Laboratory\,\cite{prex2}. Moreover, parity violating scattering 
experiments in ${}^{208}$Pb with an even higher accuracy may be 
possible in the near future at the new MESA facility in 
Mainz\,\cite{sfie13}.

\begin{figure}[t]
\includegraphics[width=0.95\linewidth,clip=true]{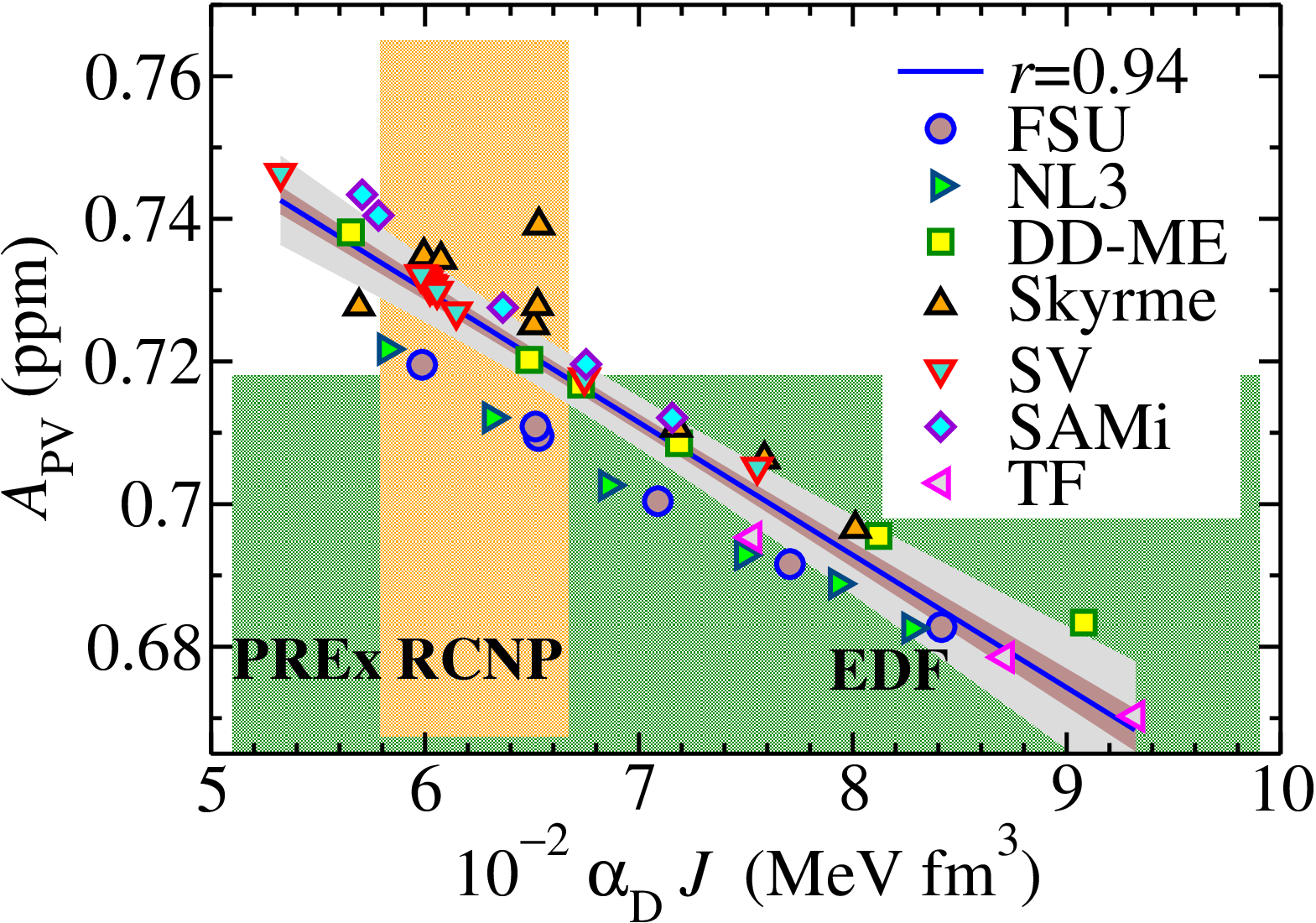}
 \caption{(Color online) 
 Parity violating asymmetry in ${}^{208}$Pb at
the PREX kinematics as a function of dipole polarizability  
times the symmetry energy at saturation predicted by
the same EDFs used in the previous figures 
\cite{skyrme1,*skyrme2,*lns,*sk255,roca-maza12b,*roca-maza13,
sv,Lalazissis:1996rd,nl3fsu1,piekarewicz11,fattoyev13,ddme2,*ddme}. The
horizontal and vertical bands correspond to the region allowed by
experimental data: $\alpha_D J = (6.23\pm 0.44)\times 10^{2}$ MeV
fm$^{3}$ and $A_{\rm PV} = 0.656\pm(0.060)_{\rm stat.} \pm
(0.014)_{\rm syst.}$ ppm. The linear fit gives $A_{\rm PV}= 0.842\pm
0.001 - (186\pm10)\times 10^{-6}\alpha_D J$ with a correlation
coefficient $r=0.94$ and the two shaded oblique regions represent the
99.9\% and 70\% confidence bands.}
 \label{fig3}
\end{figure} 

Given the strong correlation displayed by both $\alpha_D J$ and $A_{\rm PV}$ 
with the neutron skin thickness of ${}^{208}$Pb, it is natural to expect a 
close relation between $\alpha_D J$ and $A_{\rm PV}$. Note that the  parity 
violating asymmetry $A_{\rm PV}$ is the physical observable directly measured 
in the experiment.  We display in Fig.\,\ref{fig3} the predictions for $A_{\rm PV}$ 
at the PREX kinematics against $\alpha_D J$ for the same set of EDFs used in this 
work. Note that the nuclear physics input for $A_{\rm PV}$ involves both (point) 
neutron and proton densities ---as opposed to only their respective rms radii--- 
properly folded with the proton and neutron electromagnetic form factors.  
We underscore, however, that such densities are at the core of all nuclear 
density functionals, so the comparison against experiment may always be done directly 
in terms of $A_{\rm PV}$. Also shown in Fig.\,\ref{fig3} are the regions allowed 
by the experimental data in the form of a horizontal band for PREX [as given in 
Eq.(\ref{apv2})] and a vertical band for the RCNP measurement of $\alpha_D$
[$\alpha_D$=20.1$\pm$0.6 fm$^3$]---times an assumed value for the 
symmetry energy of $J\!=\![31\!\pm\!(2)_{\rm est.}]$\,MeV. That is,
\begin{equation}
  \alpha_{D}J = \big[623 \pm (19)_{\rm exp.} \pm(40)_{\rm est.}\big]
  {\rm MeV\,fm}^{3} \;.
 \label{adj}
\end{equation}
Although there is some spread in the theoretical predictions, 
the large value of $r\!=\!0.94$ suggests that the correlation between 
$\alpha_D J$ and $A_{\rm PV}$ remains strong. It is interesting to note
that a more precise measurement of $A_{\rm PV}$ in ${}^{208}$Pb with 
a central value lower than 0.7 ppm might rule out most (if not all!) of 
the state-of-the-art EDFs available in the literature. We stress 
that such a thought-provoking conclusion was reached by assuming 
a realistic range for the symmetry energy at saturation 
($29 \leq J\leq 33$\,MeV) compatible with different estimates 
\cite{tsang12,lattimer12}. 
Moreover, one may further constrain $A_{\rm PV}$ through 
its correlation with 
$\alpha_D J$. Indeed, invoking the experimental value for $\alpha_D$ 
with the alluded value for $J$ leads to:
\begin{equation}
  A_{\rm PV}\!=\!0.724\pm(0.003)_{\rm exp.}\pm(0.006)_{\rm theo.} 
 \pm(0.008)_{\rm est.}\,{\rm ppm}\,.
 \label{apv3}
\end{equation}
This would correspond to an accuracy of about 1.5\%.\\

\section{Conclusions}
\label{conclusion}
In summary, we have used insights from the droplet model
to understand correlations between the electric dipole polarizability, the 
neutron skin thickness, and the properties of the symmetry energy around
saturation density. The correlations suggested by the macroscopic droplet
model were verified in a microscopic study using a comprehensive 
set of EDFs. In particular, we found that the product of the electric dipole
polarizability $\alpha_D$ and  the symmetry energy at saturation
density $J$ is a far better isovector indicator than $\alpha_D$ alone.
We have shown that high-precision measurements of the dipole response of heavy 
nuclei (such as ${}^{208}$Pb) can significantly improve our knowledge 
of the density dependence of the symmetry energy. Indeed, the strong
correlation that we found between $\alpha_D J$ and the slope of the symmetry 
energy $L$ was used to establish a tight relation between $L$ and $J$
[see Eq.(\ref{jl})]. Moreover, by adopting the well accepted range for
the symmetry energy of
$J\!=\![31\pm (2)_{\rm est.}]$\,MeV\,\cite{tsang12,lattimer12},
the correlation between $\alpha_D J$ and the neutron skin thickness
displayed in Fig.\,\ref{fig1} suggests
$\Delta r_{np}\!\approx\! 0.168\,{\rm fm}$ for ${}^{208}$Pb,
with properly computed experimental, theoretical, and ``estimated''
uncertainties [see Eq.~(\ref{rnp0})].
Given the strong correlation between $\Delta r_{np}$ and $L$, 
we were also able to constrain the slope of the
symmetry energy at saturation density to $L\!\approx\!43\,{\rm MeV}$.
These values are consistent with the predictions for the
neutron skin thickness in ${}^{208}$Pb and $L$ extracted from
different experiments including heavy-ion collisions, giant resonances,
antiprotonic atoms, hadronic probes, and spin polarized electron
scattering (see Refs.~\cite{tsang12,lattimer12} and references
therein). They also agree nicely with the constraints on $L$ and
$\Delta r_{np}$ of ${}^{208}$Pb derived from recent astrophysical
observations supplemented with microscopic calculations of neutron
matter \cite{steiner10,hebeler10,*hebeler13}. 

Further, we found that the 
parity violating asymmetry $A_{\rm PV}$ measured by the PREX collaboration not 
only is strongly correlated with the neutron skin thickness but also with 
$\alpha_D J$. This has the advantage that theoretical calculations of 
$A_{\rm PV}$ may be directly compared against experiment---without the 
need to invoke (albeit mild) model-dependent assumptions. Ultimately, 
we have combined both observables ($A_{\rm PV}$ and $\alpha_D$) to derive 
an experimentally allowed region for the theoretical models.
The estimated uncertainties derived for $L$ and $\Delta r_{np}$ from the 
high-precision measurement of $\alpha_D$ are appreciably smaller than 
the ones expected from the 3\% measurement of $A_{\rm PV}$ at PREX-II. 
Ideally,  a 1\% accuracy on $A_{\rm PV}$ may be required to improve the 
constraint already imposed from $\alpha_D$\,\cite{roca-maza11}. 
However, we underscore that $A_{\rm PV}$ and $\alpha_D$ form 
a critical set of independent isovector indicators that could provide 
valuable insights into the nature of the nuclear density functional.

Finally, we highlighted the importance of
performing high-precision measurements of $\alpha_D$, and when
possible $A_{\rm PV}$ \cite{shufang12}, in other medium and heavy nuclei ({\sl
e.g.,} ${}^{48}$Ca, ${}^{120}$Sn, and ${}^{208}$Pb). We note 
that whereas a large number of studies---including this one---suggest 
that the neutron skin thickness of ${}^{208}$Pb is fairly thin, ruling out a 
thick neutron skin as suggested by the central value of the PREX experiment 
may be premature\,\cite{fattoyev13}. However, we are confident that systematic 
studies involving measurements of $\alpha_D$ and $A_{\rm PV}$ 
will help in constraining the density dependence of the symmetry energy.
Thus, we encourage systematic studies of these 
observables as such a program will be of enormous value in constraining 
the isovector sector of the nuclear EDF.

\begin{acknowledgments}
We are indebted to Prof. W. Nazarewicz and Prof. P.-G. Reinhard for valuable 
discussions. We also thank Prof. P.-G. Reinhard for providing us with the results 
predicted by the SV family of interactions \cite{sv}. M.C. and X.V. thank Prof. P. von Neumann-Cosel 
for useful correspondence. They acknowledge the support of the Consolider Ingenio 
2010 Programme CPAN CSD2007-00042, Grant No. FIS2011-24154
from MICINN and FEDER, and Grant No. 2009SGR-1289 from Generalitat de
Catalunya. Partial support from the US Department of Energy under 
Contract No. DE-FG05-92ER40750 (J.P.) is greatly acknowledged.
\end{acknowledgments}
%

\bibliography{bibliography}

\end{document}